\documentclass[showpacs,preprintnumbers,amsmath,amssymb]{revtex4}
\usepackage{color}
\usepackage{graphicx}
\usepackage{dcolumn}
\usepackage{multirow}
\usepackage{booktabs}
\usepackage{dsfont}
\usepackage{subfigure}
\usepackage{threeparttable}
\usepackage{bm}
\usepackage{hyperref}
\usepackage{algorithm}
\usepackage{algorithmicx}
\usepackage{algpseudocode}
\newcommand{\ie}{\emph{i.e., }}
\newcommand{\eg}{\emph{e.g., }}

\newcommand{\etal}{\emph{et al. }}

\begin{document}

\title{Fast optimal structures generator for parameterized quantum circuits}

\author{Chuangtao Chen$^2$}
\author{Zhimin He$^{1,}$}
\email{zhmihe@gmail.com}
\author{Shenggen Zheng$^{3,}$}
\email{zhengshg@pcl.ac.cn}
\author{Yan Zhou$^{1,}$}
\author{Haozhen Situ$^4$}
\affiliation{$^1$School of Electronic and Information Engineering, Foshan University, Foshan 528000, China \\
$^2$School of Mechatronic Engineering and Automation, Foshan University, Foshan 528000, China \\
$^3$Peng Cheng Laboratory, Shenzhen 518055, China\\
$^4$College of Mathematics and Informatics, South China Agricultural University, Guangzhou 510642, China \\
}%

\date{\today}

\begin{abstract}
Current structure optimization algorithms optimize the structure of quantum circuit from scratch for each new task of variational quantum algorithms (VQAs) without using any prior experience, which is inefficient and time-consuming. Besides, the number of quantum gates is a hyper-parameter of these algorithms, which is difficult and time-consuming to determine. In this paper, we propose a rapid structure optimization algorithm for VQAs which   automatically determines the number of quantum gates and directly generates the optimal structures for new tasks with the meta-trained graph variational autoencoder (VAE) on a number of training tasks. We also develop a meta-trained predictor to filter out circuits with poor performances to further accelerate the algorithm. Simulation results show that our method output structures with lower loss and it is 70$\times$ faster in running time compared to a state-of-the-art algorithm, namely DQAS.
\end{abstract}

\maketitle

\section{Introduction}
\label{intro}
Variational quantum algorithm\cite{cerezo2021variational} (VQA) is one of  the most promising strategies to achieve quantum advantage and has been  applied successfully to   problems of optimization \cite{farhi2014quantum,wang2018quantum,crooks2018performance}, quantum chemistry\cite{peruzzo2014variational,mcclean2016theory,williams2018free,heifetz2020quantum} and quantum machine learning models\cite{biamonte2017quantum,romero2017quantum,farhi2018classification,mitarai2018quantum,du2020expressive,situ2020quantum}.
The performances of VQAs  rely largely on the structures of parameterized quantum circuits.
Many VQAs use manually-designed structures, which depend heavily  on human experts and are inefficient.
Some VQAs use structure templates \cite{kandala2017hardware,hadfield2019quantum}, which are inflexible and have many redundant gates.

Structure optimization algorithms (SOAs) have been proposed to  search automatically an optimal circuit structure for a given variational quantum algorithm \cite{khatri2019quantum,li2020quantum,lu2020markovian,zhang2020differentiable,zhang2021neural,he2021variational,Moro2021quantum,ostaszewski2021reinforcement,kuo2021quantum,du2020quantum}.
However, current  SOAs search an optimal circuit structure according to a predefined circuit length. The circuit length is a hyperparameter, which is given by the prior knowledge of human experts or the simulation results of different lengths. VQA cannot converge to the optimal solution when the length of the quantum circuit is insufficient. However, there exist a lot of redundant gates resulting in many noises if the circuit length is too large.
Exploring an optimal length would be computationally expensive.

Most of  SOAs optimize the circuit structure from scratch for each new task without using any prior experience.
A MetaQAS algorithm\cite{chen2021quantum} is proposed to accelerate SOAs by using prior experience of the optimal structures in the past tasks.
However, MetaQAS only learns the initialization heuristics of the structure and gate parameters for the circuit with a fixed length. It  needs still further  optimizations on its structure and gate parameters.

To deal with  the above problems, we propose a rapid structure optimization algorithm for VQAs, which can generate directly  quantum circuits with optimal lengths and structures with respect to a given new task by a meta-trained generator.
The structure of the quantum circuit is denoted by a directed acyclic graph (DAG) \cite{nam2018automated,childs2019circuit,wu2020qgo}. We train a DAG variational autoencoder (VAE) on a variety of VQA tasks and their corresponding optimal structures. For a given new tasks, the trained VAE can  directly generate the optimal structures.
We also train  a predictor on the performances of different structures on different VQA tasks to directly filter  out the structures with poor performances.
It is  worth to  point out that the training of the generator and predictor is a one time job and can be applied to a variety of new tasks.
Simulation results show that the proposed method can generate better structures than a state-of-the-art algorithm, \ie DQAS \cite{zhang2020differentiable}. Moreover, it is 70$\times$ faster than DQAS.



\section{Method}
We propose a rapid structure optimization algorithm for parameterized quantum circuits by using the prior knowledge attained from a variety of training tasks.
The proposed method can be used to generate the optimal circuit structures for tasks of different VQAs. In this section, we illustrate how to generate the optimal quantum circuits for the tasks of variational quantum compiling, which is a typical VQA.
Quantum compiling aims to convert a target quantum circuit or unitary $ U_t $ into a native gate sequence $ U_c(\boldsymbol{\theta}) $, where $ \boldsymbol{\theta} $ is the trainable gate parameters of the compiled circuit.

\begin{figure}[h]
	\centering
	\subfigure[a quantum circuit with $5$ gates]{\includegraphics[width=5cm]{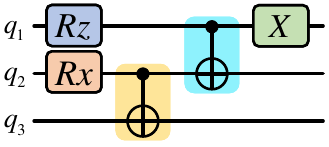}}
	\hspace{1.9cm}
	\subfigure[DAG representation of the quantum circuit]{\includegraphics[width=5cm]{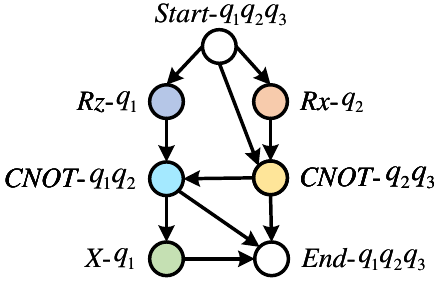}}
	\caption{An example of the DAG representation of a quantum circuit.
	{}}
	\label{fig_DAG}
\end{figure}

We use a directed acyclic graph (DAG) $ \mathcal{G}$ to represent a quantum circuit, where the nodes of the DAG denote the quantum gates and the directed edges encode their input/output relationships. A note $v$ can be denoted by $\mathit{GATE}\textendash{q_i}$, where $\mathit{GATE}$ is the type of the quantum gate and ${q_i}$ is the qubit that the gate operates on. For example, $\mathit{R_x}\textendash q_2$ denotes a rotation gate $R_x$ acting on the qubit $q_2$.
By representing an $N$-qubits quantum circuit with a DAG, we add a $\mathit{Start}\textendash{\{q_i\}_{i=1}^{N}}$ and an $\mathit{End}\textendash{\{q_i\}_{i=1}^{N}}$ nodes as the first and the last nodes of the DAG.
Fig.~\ref{fig_DAG} shows an example of the DAG representation of a quantum circuit, whose length ($L$) and depth ($D$) are $5$ and $4$.
The DAG has 7 nodes including a \emph{Start} and an \emph{End} nodes.
The DAG naturally contains the structure information of the quantum circuit such as the type of quantum gates and their connection relationships.
By using the DAG representation, we can extract the structure information of a quantum circuit with graph models\cite{hamilton2017representation,wu2020comprehensive}.

\begin{figure*}[htp]
	\centering
	\subfigure[training step]{\includegraphics[width=10cm]{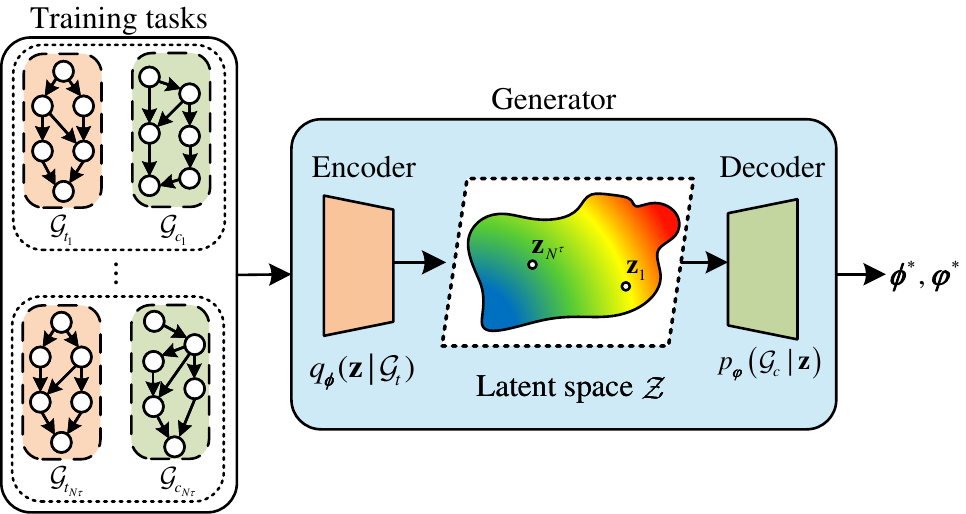}}
	\subfigure[test step]{\includegraphics[width=7cm]{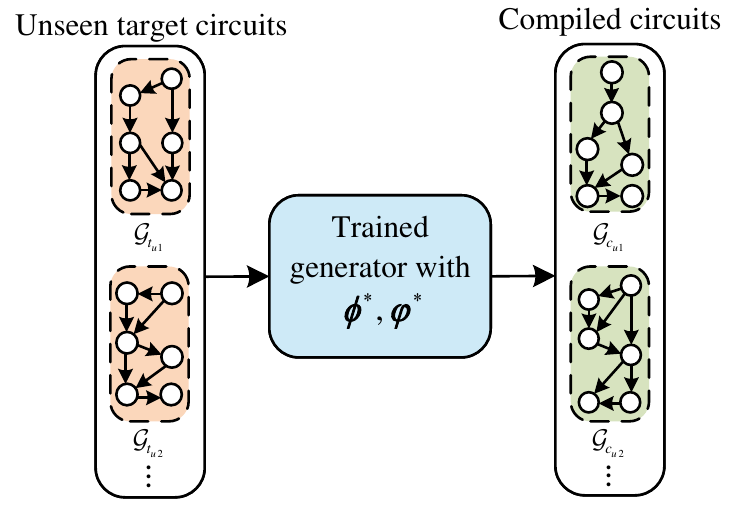}}
	\caption{The workflow of the proposed method.}
	\label{fig:trainTest}
\end{figure*}

The proposed method consists of two steps, \ie training and test steps, as shown in Fig.\,\ref{fig:trainTest}.
During the training step, we train a generator on a variety of tasks $\mathcal{D}=\{(\mathcal{G}_{t_i}, \mathcal{G}_{c_i})\}_{i=1}^{N_\tau}$ with meta-learning, where $N_\tau$ is the number of training tasks.
Each task consists of a target circuit $\mathcal{G}_{t_i}$ and its compiled circuit $\mathcal{G}_{c_i}$.
The target circuits are randomly generated and their corresponding compiled circuits can be found by any structure optimization algorithm. The generator consists of a graph encoder ($q_{\boldsymbol{\phi}}(\mathbf{z} \mid \mathcal{G}_t)$) and a graph decoder ($p_{\boldsymbol{\varphi}}(\mathcal{G}_c\mid\mathbf{z})$ ), which are used to learn a latent space ($\mathcal{Z}$) between the target circuits and the compiled ones via amortized inference, where $\boldsymbol{\phi}$ and $\boldsymbol{\varphi}$ are trainable parameters of the graph encoder and decoder.
We meta-train the generator to minimize the approximated evidence lower bound (ELBO) for each task using amortized inference
\begin{equation}
	\begin{aligned}
		\min_{\boldsymbol{\phi, \varphi}} \sum_{(\mathcal{G}_t,\mathcal{G}_c) \sim \mathcal{D}}
		(-\mathds{E}_{\mathbf{z} \sim q_{\boldsymbol{\phi}}\left(\mathbf{z} \mid \mathcal{G}_{t}\right)}(\log p_{\boldsymbol{\varphi}}\left(\mathcal{G}_{c} \mid \mathbf{z}\right))
		+\lambda \cdot \operatorname{KL}(q_{\boldsymbol{\phi}}\left(\mathbf{z} \mid \mathcal{G}_{t}\right) \| p(\mathbf{z}))).
	\end{aligned}
\end{equation}
The first term is the reconstructed loss and  the latter is Kullback-Leibler(KL) divergence between two distributions. We make $ q_{\boldsymbol{\phi}}(\mathbf{z} \mid \mathcal{G}_t)$ close to the prior distribution $ p(\mathbf{z}) $ by minimizing the KL divergence, where $ p(\mathbf{z}) $ is a standard normal distribution. $ \lambda $ is a weighted parameter which balances the reconstructed and KL loss. The optimization problem can be solved by stochastic gradient variational Bayes\cite{kingma2013auto}.
During the meta-training, we use the teacher forcing strategy \cite{jin2018junction} to calculate the reconstructed loss.

In the test phase, we can use the meta-trained generator to directly generate compiled circuits for new target circuits which are different from the training ones.
Given a new target circuit, we use the graph encoder $q_{\boldsymbol{{\phi}^{*}}}(\mathbf{z} \mid \mathcal{G}_t)$ to calculate its latent vector $ \mathbf{z}$. The decoder $p_{\boldsymbol{{\varphi}^{*}}}(\mathcal{G}_c \mid \mathbf{z})$ generates serval candidates of compiled quantum circuits based on $ \mathbf{z}$ according to Algorithm \ref{algorithm}. Then we can select the candidate with the lowest cost after fine-tuning the gate parameters as the final compiled circuit. We describe the encoder, decoder and predictor in detail in the following.


\subsection{Encoder}
\begin{figure*}
	\centering	
	\begin{minipage}[b]{0.85\textwidth}
		\centering
		\subfigure[generator]{\includegraphics[width=\textwidth]{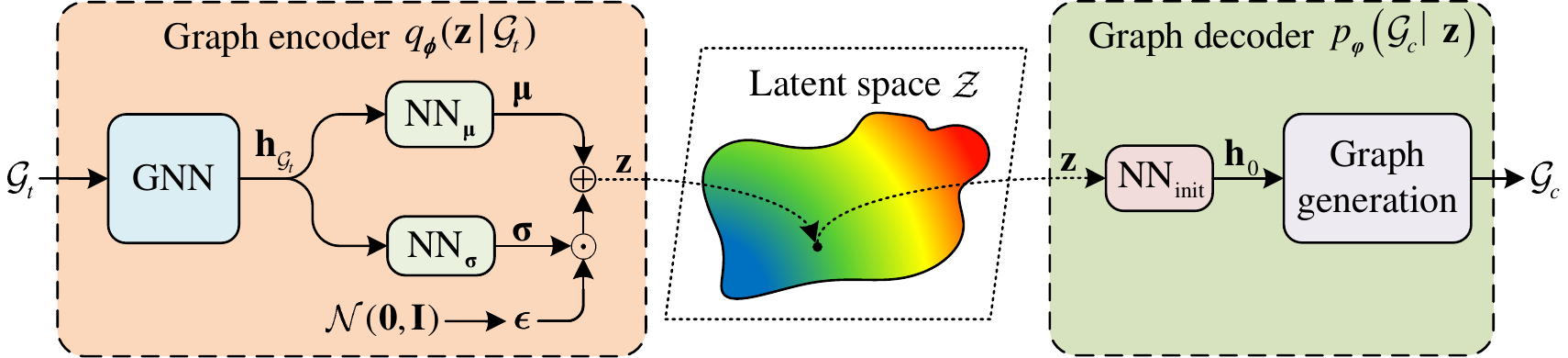}}
	\end{minipage}%
	\newline
	\begin{minipage}[b]{0.42\textwidth}
		\centering
		\subfigure[predictor]{\includegraphics[width=\textwidth]{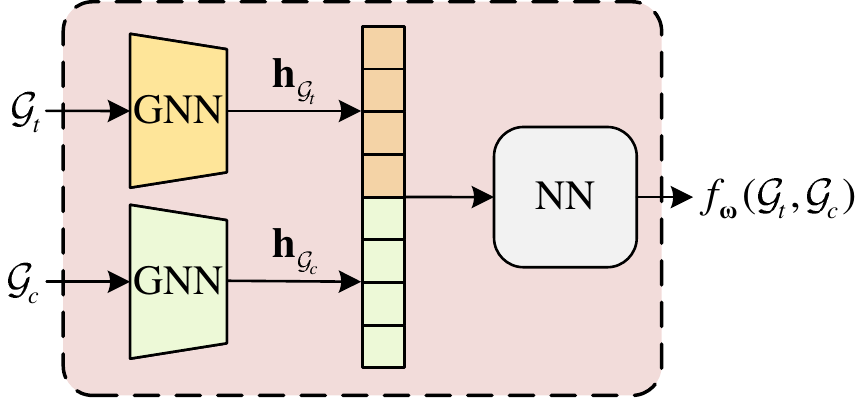}}
	\end{minipage}
	\caption{The structures of the generator and the predictor.}
	\label{fig:Generaloverview}
\end{figure*}

The encoder $q_{\boldsymbol{\phi}}(\mathbf{z} \mid \mathcal{G}_t)$ encodes the target circuit represented by a DAG  $\mathcal{G}_{t}$ into a latent code $\mathbf{z}$.
We use a graph neural network to compute the hidden state $\mathbf{h}_{v}$ of each node $v$ in $\mathcal{G}_{t}$ as shown in Fig. \ref{fig:Generaloverview}(a)
\begin{equation}
	\begin{aligned}
		\mathbf{h}_{v}=\mathcal{U}\left(\mathbf{x}_{v}, \mathbf{h}_{v}^{\text{in}}\right),\\
	\end{aligned}
	\label{eq compute hidden state}
\end{equation}
where $\mathcal{U}$ is an update function which can be implemented by a Gated Recurrent Unit (GRU) model \cite{69e088c8129341ac89810907fe6b1bfe}
\begin{equation}
	\begin{aligned}
		\mathbf{h}_{v}=\mathit{GRU}_e\left(\mathbf{x}_{v},\mathbf{h}_{v}^{\text{in}}\right).
	\end{aligned}
	\label{eq hiddenvode}
\end{equation}
$\mathbf{x}_{v}$ is a one-hot vector of the node $v$, which represents a candidate operation considering the type of the quantum gate and the qubit it acts on. $\mathbf{h}_{v}^{\mathrm{in}}=\mathcal{A}\left(\left\{\mathbf{h}_{u}: u \rightarrow v\right\}\right)$, where $\mathcal{A}$ is an aggregation function and $ u \rightarrow v$ denotes that there is a directed edge from  the node $u$ to the node $v$. $\{\mathbf{h}_{u}: u \rightarrow v\}$ is a set of hidden states of $v$'s predecessors.
The aggregation function $\mathcal{A}$ can be implemented by a gated sum function
\begin{equation}
	\begin{aligned}
		\mathbf{h}_{v}^{ \text{in}}=\sum_{u \rightarrow v} g_e\left(\mathbf{h}_{u}\right) \odot m_e\left(\mathbf{h}_{u}\right),
	\end{aligned}
	\label{eq:aggregation function}
\end{equation}
where $g_e$ is a gating network which is a single linear layer followed by a $ sigmoid $ activation function. $m_e$ is a mapping network which is a single linear layer without bias and activation function. $\odot$ is element-wise multiplication. We set $\mathbf{h}_v^{\text{in}}$ of the $\mathit{Start}$ node to be a zero vector, which has no predecessor node.
The hidden state of the $\mathit{End}$ node  is used to represent the graph state, \ie  $\mathbf{h}_{\mathcal{G}_{t}}=\mathbf{h}_{v_{End}}$.

We take $\mathbf{h}_{\mathcal{G}_t}$ as the input of two single linear layers $ \mathrm{NN}_{\boldsymbol{\mu}} $ and $ \mathrm{NN}_{\boldsymbol{\sigma}} $ to obtain the mean vector $\boldsymbol{\mu}$ and standard deviation vector $\boldsymbol{\sigma}$ as shown in Fig. \ref{fig:Generaloverview}(a).
Then the hidden vector $\mathbf{z}$ can be sampled from a circuit-conditioned Gaussian distribution, \ie  $\mathbf{z} \sim q_{\boldsymbol{\phi}}\left(\mathbf{z} \mid \mathcal{G}_{t}\right) = \mathcal{N}(\boldsymbol{\mu}, \boldsymbol{\sigma}^2)$.
By using the reparameterization trick, $\mathbf{z}$ can be denoted as $\mathbf{z}=\boldsymbol{\mu}+\boldsymbol{\sigma} \odot \boldsymbol{\epsilon}$, where $ \boldsymbol{\epsilon} \sim \mathcal{N}(\mathbf{0}, \mathbf{I}) $. The latent vector $\mathbf{z}$ is an embedding of $ \mathcal{G}_t $ in the latent space $ \mathcal{Z} $, which is the input of the decoder.

\subsection{Decoder}
\label{decoder}
Given a latent vector, the graph decoder $p_{\boldsymbol{{\varphi}}}(\mathcal{G} \mid \mathbf{z})$ outputs a DAG $\mathcal{G}$ which represents a quantum circuit.
The detailed process is shown in Algorithm \ref{algorithm}. A single linear layer $ \mathrm{NN}_{init} $ with $\mathit{tanh}$ activity function is used to calculate the hidden state of the  $\mathit{Start}$ node $v_0$. Then the hidden state of the subsequent node $v_i$ is calculated by $\mathit{GRU}_d $, $ g_d $ and $ m_d$, which have the same structures with $\mathit{GRU}_e $, $ g_e $ and $ m_e$ used in the encoder.
We progressively sample and add a new node $v_{i+1}$ to $\mathcal{G}$ based on the probability distribution determined by the hidden state of its last note $v_i$, \ie $\boldsymbol{p}=f_{p}(\mathbf{h}_{v_i})$, until an $End$  node is sampled or the circuit length reaches the defined maximum length, where $ f_{p} $ is a single linear layer followed by a $ \mathit{softmax} $ function.

\begin{algorithm}
	\renewcommand{\algorithmicrequire}{\textbf{Input:}}
	\renewcommand\algorithmicensure {\textbf{Output:} }
	\caption{DAG generation with a decoder.}
	\label{algorithm}
	\begin{algorithmic}[1]
		\Require $\mathbf{z}$: a latent vector.
		\State $\mathcal{G}$ =\{\}, $i=0$ and add a $Start$ node to $\mathcal{G}$.
		\State calculate the hidden state of the $Start$ node $\mathbf{h}_{0} = \mathrm{NN}_{init}(\mathbf{z})$.
		\Repeat
		\State calculate the hidden state $\mathbf{h}_{v_{i}} $ of the node $ v_{i} $ using $ \mathit{GRU}_d $, $ g_d $ and $ m_d $.
		\State sample and add the next node $v_{i+1}$ to $\mathcal{G}$ based on the probability distribution $\boldsymbol{p}=f_{p}(\mathbf{h}_{v_i})$.
		\State add directed edge to $v_{i+1}$ according to the operated qubits of the selected gate.
		\State $i=i+1$
		\Until{$v_{i}$ is an $End$ node or the circuit length reaches the maximum length}
	\end{algorithmic}
\end{algorithm}

\subsection{Predictor}
Current SOAs require to evaluate the performances of a large number of circuit structures on NISQ devises, which requires a large amount of running time.
Zhang \etal trained a neural predictor with a small subset of quantum circuits to estimate the performances of different structures\cite{zhang2021neural}.
The predictor is trained for a specific task and requires to retrain for a new task.

A meta-trained predictor on a variety of tasks $\mathcal{D}_{pre}$ can directly predict the performances of quantum circuits for different unseen tasks instead of getting the loss on the quantum devices after optimizing its gate parameters, which can significantly reduce the evaluation time of circuit structures. Each training sample consists of a target circuit, a compiled circuit and the corresponding loss, which are represented by $ \mathcal{G}_{t} $, $ \mathcal{G}_{c} $ and $s$. The structure of the predictor is described in Fig.\,\ref{fig:Generaloverview}(b).
The DAGs $ \mathcal{G}_{t} $ and $ \mathcal{G}_{c} $  are first converted to hidden vectors $\mathbf{h}_{\mathcal{G}_t}$ and $\mathbf{h}_{\mathcal{G}_c}$ by two graph neural networks, whose structures are the same as the one used in the encoder.
$\mathbf{h}_{\mathcal{G}_t}$ and $\mathbf{h}_{\mathcal{G}_c} $ are feeded into two linear layers with $ \mathit{relu} $ to get the predicted loss $f_{\boldsymbol{\omega}}(\mathcal{G}_{t}, \mathcal{G}_{c})$, where $\boldsymbol{\omega}$ are the trainable parameters of the predictor.
The predictor are trained by minimizing the Mean-Squared Loss between the predicted loss and the true one
\begin{equation}
	\begin{aligned}
		\min_{\boldsymbol{\omega}}\sum_{(\mathcal{G}_{t}, \mathcal{G}_{c}, s) \sim \mathcal{D}_{pre}}\left(s-f_{\boldsymbol{\omega}}(\mathcal{G}_{t}, \mathcal{G}_{c})\right)^{2}.
	\end{aligned}
	\label{eq:predictor loss function}
\end{equation}

\section{Numerical simulation}
In this section, we show the simulation results of the meta-trained generator and predictor on variational quantum compiling, and compare the performance of the proposed method to a state-of-the-art structure optimization algorithm, \ie DQAS \cite{zhang2020differentiable}. All the simulations are run on a  classical computer with a CPU i9-10900K.
We consider target circuits with 3 qubits. The target circuits consist of 4, 5 or 6 quantum gates which are randomly selected from the gate set $\mathcal{A}_{\mathit{target}}=$\{$\mathit{H}$, $\mathit{PauliX}$, $\mathit{PauliY}$, $\mathit{PauliZ}$, $\mathit{S}$, $\mathit{T}$, $\mathit{R_X}(\theta)$, $\mathit{R_Y}(\theta)$, $\mathit{R_Z}(\theta)$, $\mathit{CNOT}$, $\mathit{CZ}$, $\mathit{CY}$, $\mathit{SWAP}$, $\mathit{Toffoli}$, $\mathit{CSWAP}$\} and act on randomly selected qubits.
The training dataset of the generator contains 3000 samples and each sample consists of the DAGs of the target and the compiled circuits.
The native gate set for variational quantum compiling is $\mathcal{A}_{\mathit{native}}=$\{$\mathit{RX}(\pi, \pm\pi/2)$, $\mathit{RZ}(\theta)$, $\mathit{CRZ}(\theta)$, $\mathit{CZ}$, $\mathit{XY}(\theta)$\}, which is used by Rigetti's Aspen-11 quantum processor \cite{Aspen11QPU}.
We set the maximum length of the compiled circuit to 30.
The generator is trained until the loss converges, which takes 1.57 hours.

After training, we evaluate the generator with 300 target circuits which do not exist in the training dataset.
For each test circuit, the generator generates 100 structures of the compiled circuit. We get the LHST loss \cite{khatri2019quantum} of each structure after its gate parameters are optimized, and output the one with the lowest loss.
As described in Section \ref{decoder}, the graph decoder generates a circuit by sequentially choosing candidate gates from the native gate set based on the probability distribution of candidate gates.
We consider different sampling strategies , including stochastic and top-$k$ sampling schemes \cite{fan2018hierarchical}, where $k=10, 15, 20, 25$. Stochastic and top-$k$ sampling means that the decoder selects candidate gates according to the probability distribution of all the candidate gates and the $k$ candidate gates with the highest probabilities.


We show  the average loss of the proposed method in Tabel \ref{tab:generated result all connection all 100}.
We also show the length ($L$) and the depth ($D$) of the compiled circuits, which are defined as the number of quantum gates and the number of layers, respectively.
The stochastic scheme achieves the lowest loss. However, its compiled circuits use more quantum gates and have a larger depth.
For the top-$k$ scheme, the loss decreases with the increasing of $k$. More gates are used in the compiled circuit for a larger $k$. However, the depths of compiled circuits with different $k$ are similar. The top-$k$ scheme makes a balance between the loss and the depth of the circuit. In the following simulations, we use top-$25$ scheme.
\begin{table}
	\centering
	\caption{The average loss of compiling 300 target circuits with the proposed generator.}
	\begin{tabular}{cccccc}
		\hline
		Strategy & Loss & $ L $ & $ D $ & Uniqueness (\%) & Novelty (\%) \\
		\hline
		Top-10 & 0.0333 & 14.94 & 12.30 & 99.96 & 100.00 \\
		Top-15 & 0.0205 & 16.33 & 12.73 & 99.99 & 100.00   \\
		Top-20 & 0.0153 & 16.50 & 12.42 & 99.99 & 100.00   \\
		Top-25 & 0.0133 & 17.01 & 12.28 & 99.99 & 100.00  \\
		Stochastic  & 0.0071 & 24.78 & 16.09 & 97.71 & 100.00  \\
		\hline
	\end{tabular}%
	\label{tab:generated result all connection all 100}%
\end{table}%


We also show the uniqueness and the novelty of the generated circuits in Table \ref{tab:generated result all connection all 100}, which are defined as the percentage of unique circuits in the generated circuits and the percentage of circuits that do not exist in the training set.
The quantum circuits generated by different sampling schemes have a very low repeatability and the novelty is 100.00\%, which demonstrates the good capability of the proposed model to generate optimal quantum circuits for different tasks, rather than just simply copying the circuits from the training set.

In previous simulations, we assumed that the qubits are fully connected. However, the qubits are not fully connected in the NISQ era.
We consider the chain connection, $ i.e. $, $ {q_1}\textendash{q_2}\textendash{q_3} $ and use the meta-trained generator to generate circuits under limited connections.
The search space in the chain connection is a subspace of the fully connected one.
There is no need to recollect training data and retrain the generator. Instead, we can simply add a mask code which forces the generator to only choose permitted operations under limited connections. 
The simulation results on chain connection is shown in Table \ref{tab:generated result limited connection}.
The average losses under limited connections are similar to those under fully connections. As the connection between ${q_1}$ and ${q_3}$ is forbidden, it requires more gates and larger depth.
\begin{table}[h]
	\centering
	\caption{The average losses of compiling 300 target circuits with the proposed generator under limited connections.}
	\begin{tabular}{cccccc}
    \hline
		\toprule
		Strategy & Loss & $ L $ & $ D $  \\
    \hline
		\midrule

		Top-10 & 0.0316 & 15.46 & 12.73 \\
		Top-15 & 0.0201 & 16.48 & 13.07 \\
		Top-20 & 0.0155 & 17.46 & 12.99\\
		Top-25 & 0.0144 & 17.71 & 12.96 \\
\hline
		Stochastic  & 0.0070 & 25.49 & 16.42  \\
\hline
		\bottomrule
	\end{tabular}%
	\label{tab:generated result limited connection}%
\end{table}%

\begin{figure}[h]
	\centering
	\includegraphics[width=7.2cm]{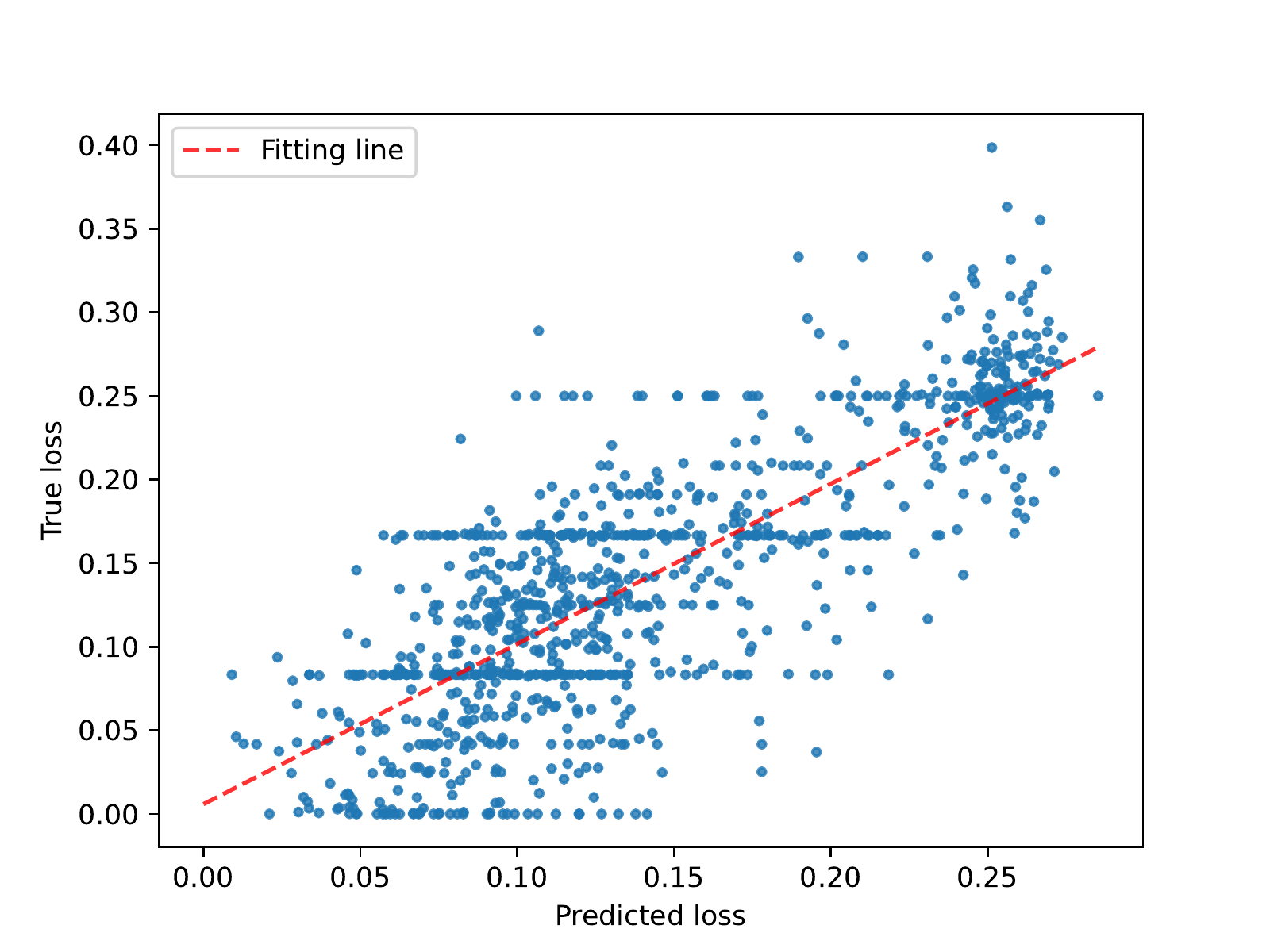}\\
	\caption{The relationship between the predicted loss and the ground-truth one.}
	\label{fig_true}
\end{figure}

\begin{figure}[h]
	\centering
	\includegraphics[width=7.3cm]{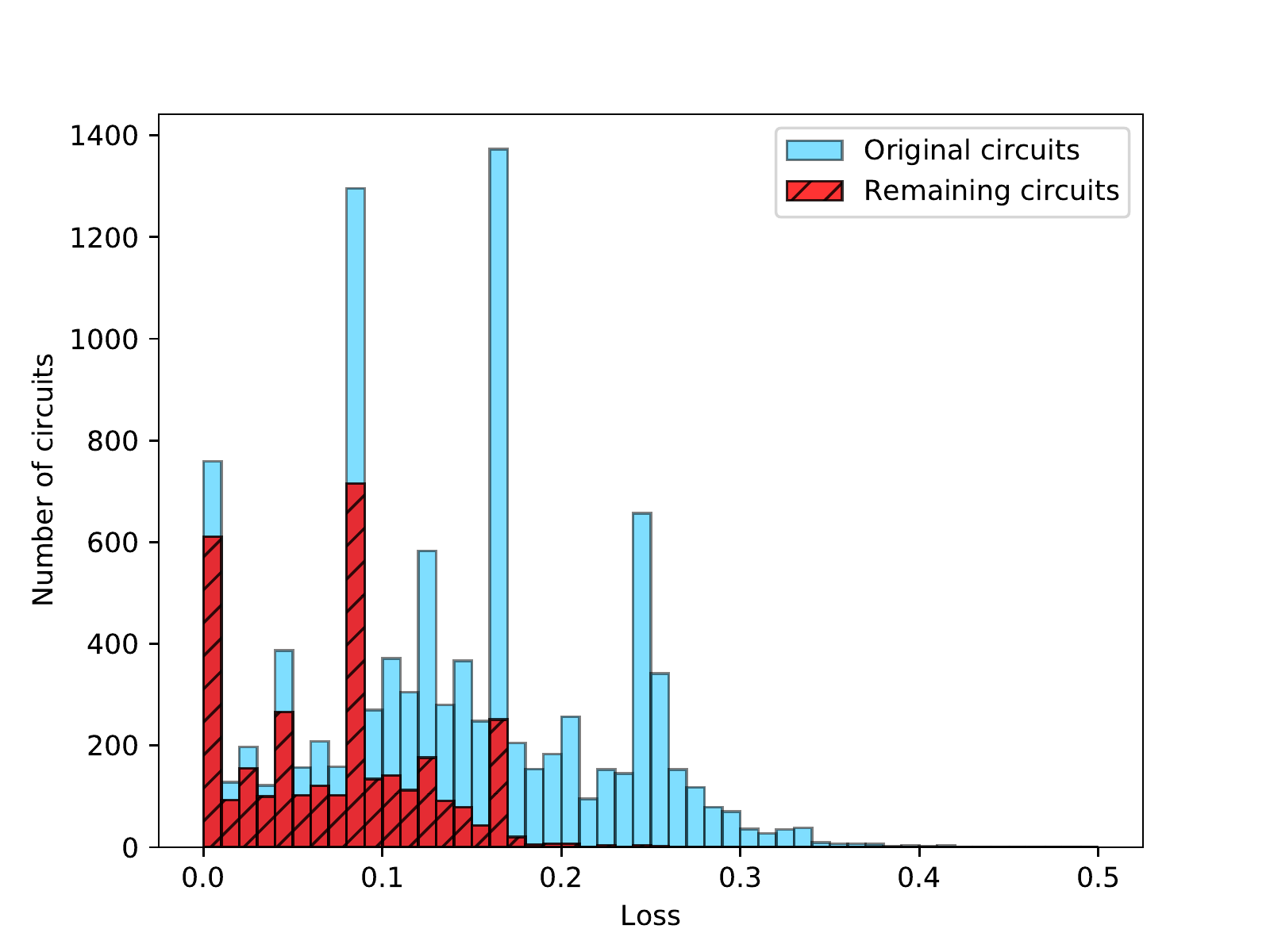}\\
	\caption{The loss distributions of 10,000 test samples (blue bars) and the remaining samples filtered by the predictor (red bars).}
	\label{fig:predicted_result1}
\end{figure}

We train a predictor to accelerate the proposed method by removing the generated circuits with unsatisfying performance.
The training dataset consists of 200,000 samples and each sample contains the DAGs of a target and its compiled circuit as well as the corresponding loss.
Please see Appendix \ref{sec:AppendixE} for more details of the training and test dateset.
The predictor is trained for 100 epochs. The Pearson correlation coefficient between the predicted and the true loss on 10,000 test samples is 0.784, which indicates a strong correlation between the predicted loss and the true one.
For ease of visualization, we illustrate the predicted and true losses of 1,000 randomly selected test samples in Fig.~\ref{fig_true}.
We show the loss distributions of 10,000 test samples and the remaining samples after removing the ones whose predicted losses are larger than 0.1 in Fig. \ref{fig:predicted_result1}.
We can observe that the predictor can filter out most of compiled circuits with poor performances while retain circuits with high performances.


We compare the performance of the proposed method with a state-of-the-art structure optimization algorithm, namely DQAS \cite{zhang2020differentiable}, in terms of loss and running time. As the circuit length $L$ is a hyperparameter of DQAS, we gradually increase $L$ until the loss is less than 0.05 or the circuit length reaches 30.
The simulation results are shown in the Table \ref{tab:Comparison of search results without predictor} and Table \ref{tab:Comparison of search results time}.
The proposed methods achieve lower losses than DQAS. The average length and depth of the proposed methods with top-25 scheme are similar to the ones of DQAS.
By using stochastic scheme, the proposed method achieves the lowest loss, \ie 0.0075. However, it uses 8  more gates, leading to a 4 increase in depth. The loss, length and depth of the proposed method with a predictor is similar to the one without predictor.
\begin{table}[h]
	\centering
	\caption{Simulation results of different structure optimization algorithms. ``gen'' and ``pred'' denote the generator and predictor, respectively.}
	\begin{tabular}{cccc}
		\hline
		Method & Loss  & $ L $ & $ D $ \\
		\hline
		DQAS\cite{zhang2020differentiable}  & 0.0269 & 17.44 & \bf{12.27} \\
		\hline
		Gen-top-25 & 0.0141 & \bf{17.02} & 12.36 \\
		Gen-stochastic  & \bf{0.0075} & 25.01 & 16.39  \\
		\hline
		Gen-pred-top-25 & 0.0150 & 17.69 & 12.92\\
		Gen-pred-stochastic  & 0.0086 & 26.81 & 17.54\\
		\hline
	\end{tabular}%
	\label{tab:Comparison of search results without predictor}%
\end{table}%

Tabel \ref{tab:Comparison of search results time} shows the running time of different structure optimization algorithms.
We define the times of searching the optimal structures and fine-tuning the gate parameters as $t_{s}$ and $t_{f}$, respectively.
The running time of the predictor is denoted as $t_{p}$.  $t_{total}$ shows the total running time of each algorithm.
DQAS takes 8.4 hours to search the optimal structure while the proposed method only need about 1 second to generate compiled circuits.
The running time of fine-tuning step in our method is higher than DQAS as we need to fine-tune the gate parameters of a set of generated circuits.
The total running times of the proposed method with top-25 and stochastic schemes are 13.35 and 15.15 minutes, which are only 2.6\% and 3.0\% of that used by DQAS.
The predictor takes less than haft a second to predict the performances of different generated circuits. By filtering out the generated circuits with poor performances, we can largely reduce the running times in the fine-tuning step, which reduces the total running time by half.

\begin{table}[h]
	\centering
	\caption{Running times of compiling 90 target circuits with different structure optimization algorithms.}
	\begin{tabular}{cccccc}
		\hline
		Method &  $t_{s}$  & $t_{p}$ & $t_{f}$ &  $t_{total}$ \\
		\hline
		DQAS\cite{zhang2020differentiable} & 8.36 (h) &  /  & 2.66(min) & 8.40(h) \\
		\hline
		Gen-top-25 & 0.98(s) & / & {13.31(min)} & {13.35(min)} \\
		Gen-stochastic  & 1.05(s) & / & {15.13(min)} & {15.15(min)} \\
		\hline
		Gen-Pred-top-25 & 0.98(s) & 0.36(s) & 7.08(min) & 7.10(min) \\
		Gen-Pred-stochastic  & 1.05(s) & 0.40(s) & 6.95(min) & 6.98(min) \\
		\hline
	\end{tabular}%
	\label{tab:Comparison of search results time}%
\end{table}%


\section{Conclusion}
In this paper, we proposed to  generate directly optimal circuits for new tasks with a meta-trained graph variational autoencoder on a variety of training tasks.
Numerical experimental results on variational quantum compiling showed  that the proposed algorithm can generate quantum circuits with similar performance as the DQAS algorithm while taking only 2.6\% of its running time. We also proposed a predictor to filter out the structures with unsatisfying performances, which can reduce the total running time by 50\%.
We will apply the proposed algorithm to other VQAs, \eg Variational Quantum Eigensolver (VQE) and Quantum Approximate Optimization Algorithm (QAOA) in future works.

\begin{acknowledgments}
This work is supported by Guangdong Basic and Applied Basic Research Foundation (Nos.\,2021A1515012138, 2019A1515011166, 2020A1515011204), Guangdong Provincial Education Department Key platform and Scientific Research Projects (No.\,2020KTSCX132) and Key Research Project of Guangdong Province (No.\,2019KZDXM007).
\end{acknowledgments}

\bibliography{bibDB}

\begin{thebibliography}{40}
\expandafter\ifx\csname natexlab\endcsname\relax\def\natexlab#1{#1}\fi
\expandafter\ifx\csname bibnamefont\endcsname\relax
  \def\bibnamefont#1{#1}\fi
\expandafter\ifx\csname bibfnamefont\endcsname\relax
  \def\bibfnamefont#1{#1}\fi
\expandafter\ifx\csname citenamefont\endcsname\relax
  \def\citenamefont#1{#1}\fi
\expandafter\ifx\csname url\endcsname\relax
  \def\url#1{\texttt{#1}}\fi
\expandafter\ifx\csname urlprefix\endcsname\relax\def\urlprefix{URL }\fi
\providecommand{\bibinfo}[2]{#2}
\providecommand{\eprint}[2][]{\url{#2}}

\bibitem[{\citenamefont{Cerezo et~al.}(2021)\citenamefont{Cerezo, Arrasmith,
  Babbush, Benjamin, Endo, Fujii, McClean, Mitarai, Yuan, Cincio
  et~al.}}]{cerezo2021variational}
\bibinfo{author}{\bibfnamefont{M.}~\bibnamefont{Cerezo}},
  \bibinfo{author}{\bibfnamefont{A.}~\bibnamefont{Arrasmith}},
  \bibinfo{author}{\bibfnamefont{R.}~\bibnamefont{Babbush}},
  \bibinfo{author}{\bibfnamefont{S.~C.} \bibnamefont{Benjamin}},
  \bibinfo{author}{\bibfnamefont{S.}~\bibnamefont{Endo}},
  \bibinfo{author}{\bibfnamefont{K.}~\bibnamefont{Fujii}},
  \bibinfo{author}{\bibfnamefont{J.~R.} \bibnamefont{McClean}},
  \bibinfo{author}{\bibfnamefont{K.}~\bibnamefont{Mitarai}},
  \bibinfo{author}{\bibfnamefont{X.}~\bibnamefont{Yuan}},
  \bibinfo{author}{\bibfnamefont{L.}~\bibnamefont{Cincio}},
  \bibnamefont{et~al.}, \bibinfo{journal}{Nat. Rev. Phys.} pp.
  \bibinfo{pages}{1--20} (\bibinfo{year}{2021}).

\bibitem[{\citenamefont{Farhi et~al.}(2014)\citenamefont{Farhi, Goldstone, and
  Gutmann}}]{farhi2014quantum}
\bibinfo{author}{\bibfnamefont{E.}~\bibnamefont{Farhi}},
  \bibinfo{author}{\bibfnamefont{J.}~\bibnamefont{Goldstone}},
  \bibnamefont{and} \bibinfo{author}{\bibfnamefont{S.}~\bibnamefont{Gutmann}},
  \bibinfo{journal}{arXiv:1411.4028}  (\bibinfo{year}{2014}).

\bibitem[{\citenamefont{Wang et~al.}(2018)\citenamefont{Wang, Hadfield, Jiang,
  and Rieffel}}]{wang2018quantum}
\bibinfo{author}{\bibfnamefont{Z.}~\bibnamefont{Wang}},
  \bibinfo{author}{\bibfnamefont{S.}~\bibnamefont{Hadfield}},
  \bibinfo{author}{\bibfnamefont{Z.}~\bibnamefont{Jiang}}, \bibnamefont{and}
  \bibinfo{author}{\bibfnamefont{E.~G.} \bibnamefont{Rieffel}},
  \bibinfo{journal}{Phys. Rev. A} \textbf{\bibinfo{volume}{97}},
  \bibinfo{pages}{022304} (\bibinfo{year}{2018}).

\bibitem[{\citenamefont{Crooks}(2018)}]{crooks2018performance}
\bibinfo{author}{\bibfnamefont{G.~E.} \bibnamefont{Crooks}},
  \bibinfo{journal}{arXiv:1811.08419}  (\bibinfo{year}{2018}).

\bibitem[{\citenamefont{Peruzzo et~al.}(2014)\citenamefont{Peruzzo, McClean,
  Shadbolt, Yung, Zhou, Love, Aspuru-Guzik, and
  O'brien}}]{peruzzo2014variational}
\bibinfo{author}{\bibfnamefont{A.}~\bibnamefont{Peruzzo}},
  \bibinfo{author}{\bibfnamefont{J.}~\bibnamefont{McClean}},
  \bibinfo{author}{\bibfnamefont{P.}~\bibnamefont{Shadbolt}},
  \bibinfo{author}{\bibfnamefont{M.-H.} \bibnamefont{Yung}},
  \bibinfo{author}{\bibfnamefont{X.-Q.} \bibnamefont{Zhou}},
  \bibinfo{author}{\bibfnamefont{P.~J.} \bibnamefont{Love}},
  \bibinfo{author}{\bibfnamefont{A.}~\bibnamefont{Aspuru-Guzik}},
  \bibnamefont{and} \bibinfo{author}{\bibfnamefont{J.~L.}
  \bibnamefont{O'brien}}, \bibinfo{journal}{Nat. Commun.}
  \textbf{\bibinfo{volume}{5}}, \bibinfo{pages}{1} (\bibinfo{year}{2014}).

\bibitem[{\citenamefont{McClean et~al.}(2016)\citenamefont{McClean, Romero,
  Babbush, and Aspuru-Guzik}}]{mcclean2016theory}
\bibinfo{author}{\bibfnamefont{J.~R.} \bibnamefont{McClean}},
  \bibinfo{author}{\bibfnamefont{J.}~\bibnamefont{Romero}},
  \bibinfo{author}{\bibfnamefont{R.}~\bibnamefont{Babbush}}, \bibnamefont{and}
  \bibinfo{author}{\bibfnamefont{A.}~\bibnamefont{Aspuru-Guzik}},
  \bibinfo{journal}{New J. Phys.} \textbf{\bibinfo{volume}{18}},
  \bibinfo{pages}{023023} (\bibinfo{year}{2016}).

\bibitem[{\citenamefont{Williams-Noonan
  et~al.}(2018)\citenamefont{Williams-Noonan, Yuriev, and
  Chalmers}}]{williams2018free}
\bibinfo{author}{\bibfnamefont{B.~J.} \bibnamefont{Williams-Noonan}},
  \bibinfo{author}{\bibfnamefont{E.}~\bibnamefont{Yuriev}}, \bibnamefont{and}
  \bibinfo{author}{\bibfnamefont{D.~K.} \bibnamefont{Chalmers}},
  \bibinfo{journal}{J. Med. Chem.} \textbf{\bibinfo{volume}{61}},
  \bibinfo{pages}{638} (\bibinfo{year}{2018}).

\bibitem[{\citenamefont{Heifetz}(2020)}]{heifetz2020quantum}
\bibinfo{author}{\bibfnamefont{A.}~\bibnamefont{Heifetz}},
  \emph{\bibinfo{title}{Quantum Mechanics In Drug Discovery}}
  (\bibinfo{publisher}{Springer}, \bibinfo{year}{2020}).

\bibitem[{\citenamefont{Biamonte et~al.}(2017)\citenamefont{Biamonte, Wittek,
  Pancotti, Rebentrost, Wiebe, and Lloyd}}]{biamonte2017quantum}
\bibinfo{author}{\bibfnamefont{J.}~\bibnamefont{Biamonte}},
  \bibinfo{author}{\bibfnamefont{P.}~\bibnamefont{Wittek}},
  \bibinfo{author}{\bibfnamefont{N.}~\bibnamefont{Pancotti}},
  \bibinfo{author}{\bibfnamefont{P.}~\bibnamefont{Rebentrost}},
  \bibinfo{author}{\bibfnamefont{N.}~\bibnamefont{Wiebe}}, \bibnamefont{and}
  \bibinfo{author}{\bibfnamefont{S.}~\bibnamefont{Lloyd}},
  \bibinfo{journal}{Nature} \textbf{\bibinfo{volume}{549}},
  \bibinfo{pages}{195} (\bibinfo{year}{2017}).

\bibitem[{\citenamefont{Romero et~al.}(2017)\citenamefont{Romero, Olson, and
  Aspuru-Guzik}}]{romero2017quantum}
\bibinfo{author}{\bibfnamefont{J.}~\bibnamefont{Romero}},
  \bibinfo{author}{\bibfnamefont{J.~P.} \bibnamefont{Olson}}, \bibnamefont{and}
  \bibinfo{author}{\bibfnamefont{A.}~\bibnamefont{Aspuru-Guzik}},
  \bibinfo{journal}{Quantum Sci. Technol.} \textbf{\bibinfo{volume}{2}},
  \bibinfo{pages}{045001} (\bibinfo{year}{2017}).

\bibitem[{\citenamefont{Farhi and Neven}(2018)}]{farhi2018classification}
\bibinfo{author}{\bibfnamefont{E.}~\bibnamefont{Farhi}} \bibnamefont{and}
  \bibinfo{author}{\bibfnamefont{H.}~\bibnamefont{Neven}},
  \bibinfo{journal}{arXiv:1802.06002}  (\bibinfo{year}{2018}).

\bibitem[{\citenamefont{Mitarai et~al.}(2018)\citenamefont{Mitarai, Negoro,
  Kitagawa, and Fujii}}]{mitarai2018quantum}
\bibinfo{author}{\bibfnamefont{K.}~\bibnamefont{Mitarai}},
  \bibinfo{author}{\bibfnamefont{M.}~\bibnamefont{Negoro}},
  \bibinfo{author}{\bibfnamefont{M.}~\bibnamefont{Kitagawa}}, \bibnamefont{and}
  \bibinfo{author}{\bibfnamefont{K.}~\bibnamefont{Fujii}},
  \bibinfo{journal}{Phys. Rev. A} \textbf{\bibinfo{volume}{98}},
  \bibinfo{pages}{032309} (\bibinfo{year}{2018}).

\bibitem[{\citenamefont{Du et~al.}(2020{\natexlab{a}})\citenamefont{Du, Hsieh,
  Liu, and Tao}}]{du2020expressive}
\bibinfo{author}{\bibfnamefont{Y.}~\bibnamefont{Du}},
  \bibinfo{author}{\bibfnamefont{M.-H.} \bibnamefont{Hsieh}},
  \bibinfo{author}{\bibfnamefont{T.}~\bibnamefont{Liu}}, \bibnamefont{and}
  \bibinfo{author}{\bibfnamefont{D.}~\bibnamefont{Tao}},
  \bibinfo{journal}{Phys. Rev. Research} \textbf{\bibinfo{volume}{2}},
  \bibinfo{pages}{033125} (\bibinfo{year}{2020}{\natexlab{a}}).

\bibitem[{\citenamefont{Situ et~al.}(2020)\citenamefont{Situ, He, Wang, Li, and
  Zheng}}]{situ2020quantum}
\bibinfo{author}{\bibfnamefont{H.}~\bibnamefont{Situ}},
  \bibinfo{author}{\bibfnamefont{Z.}~\bibnamefont{He}},
  \bibinfo{author}{\bibfnamefont{Y.}~\bibnamefont{Wang}},
  \bibinfo{author}{\bibfnamefont{L.}~\bibnamefont{Li}}, \bibnamefont{and}
  \bibinfo{author}{\bibfnamefont{S.}~\bibnamefont{Zheng}},
  \bibinfo{journal}{Inform. Sciences} \textbf{\bibinfo{volume}{538}},
  \bibinfo{pages}{193} (\bibinfo{year}{2020}).

\bibitem[{\citenamefont{Kandala et~al.}(2017)\citenamefont{Kandala, Mezzacapo,
  Temme, Takita, Brink, Chow, and Gambetta}}]{kandala2017hardware}
\bibinfo{author}{\bibfnamefont{A.}~\bibnamefont{Kandala}},
  \bibinfo{author}{\bibfnamefont{A.}~\bibnamefont{Mezzacapo}},
  \bibinfo{author}{\bibfnamefont{K.}~\bibnamefont{Temme}},
  \bibinfo{author}{\bibfnamefont{M.}~\bibnamefont{Takita}},
  \bibinfo{author}{\bibfnamefont{M.}~\bibnamefont{Brink}},
  \bibinfo{author}{\bibfnamefont{J.~M.} \bibnamefont{Chow}}, \bibnamefont{and}
  \bibinfo{author}{\bibfnamefont{J.~M.} \bibnamefont{Gambetta}},
  \bibinfo{journal}{Nature} \textbf{\bibinfo{volume}{549}},
  \bibinfo{pages}{242} (\bibinfo{year}{2017}).

\bibitem[{\citenamefont{Hadfield et~al.}(2019)\citenamefont{Hadfield, Wang,
  O’Gorman, Rieffel, Venturelli, and Biswas}}]{hadfield2019quantum}
\bibinfo{author}{\bibfnamefont{S.}~\bibnamefont{Hadfield}},
  \bibinfo{author}{\bibfnamefont{Z.}~\bibnamefont{Wang}},
  \bibinfo{author}{\bibfnamefont{B.}~\bibnamefont{O’Gorman}},
  \bibinfo{author}{\bibfnamefont{E.~G.} \bibnamefont{Rieffel}},
  \bibinfo{author}{\bibfnamefont{D.}~\bibnamefont{Venturelli}},
  \bibnamefont{and} \bibinfo{author}{\bibfnamefont{R.}~\bibnamefont{Biswas}},
  \bibinfo{journal}{Algorithms} \textbf{\bibinfo{volume}{12}},
  \bibinfo{pages}{34} (\bibinfo{year}{2019}).

\bibitem[{\citenamefont{Khatri et~al.}(2019)\citenamefont{Khatri, LaRose,
  Poremba, Cincio, Sornborger, and Coles}}]{khatri2019quantum}
\bibinfo{author}{\bibfnamefont{S.}~\bibnamefont{Khatri}},
  \bibinfo{author}{\bibfnamefont{R.}~\bibnamefont{LaRose}},
  \bibinfo{author}{\bibfnamefont{A.}~\bibnamefont{Poremba}},
  \bibinfo{author}{\bibfnamefont{L.}~\bibnamefont{Cincio}},
  \bibinfo{author}{\bibfnamefont{A.~T.} \bibnamefont{Sornborger}},
  \bibnamefont{and} \bibinfo{author}{\bibfnamefont{P.~J.} \bibnamefont{Coles}},
  \bibinfo{journal}{Quantum} \textbf{\bibinfo{volume}{3}}, \bibinfo{pages}{140}
  (\bibinfo{year}{2019}).

\bibitem[{\citenamefont{Li et~al.}(2020)\citenamefont{Li, Fan, Coram, Riley,
  Leichenauer et~al.}}]{li2020quantum}
\bibinfo{author}{\bibfnamefont{L.}~\bibnamefont{Li}},
  \bibinfo{author}{\bibfnamefont{M.}~\bibnamefont{Fan}},
  \bibinfo{author}{\bibfnamefont{M.}~\bibnamefont{Coram}},
  \bibinfo{author}{\bibfnamefont{P.}~\bibnamefont{Riley}},
  \bibinfo{author}{\bibfnamefont{S.}~\bibnamefont{Leichenauer}},
  \bibnamefont{et~al.}, \bibinfo{journal}{Phys. Rev. Research}
  \textbf{\bibinfo{volume}{2}}, \bibinfo{pages}{023074} (\bibinfo{year}{2020}).

\bibitem[{\citenamefont{Lu et~al.}(2021)\citenamefont{Lu, Shen, and
  Deng}}]{lu2020markovian}
\bibinfo{author}{\bibfnamefont{Z.}~\bibnamefont{Lu}},
  \bibinfo{author}{\bibfnamefont{P.-X.} \bibnamefont{Shen}}, \bibnamefont{and}
  \bibinfo{author}{\bibfnamefont{D.-L.} \bibnamefont{Deng}},
  \bibinfo{journal}{Phys. Rev. Appl.} \textbf{\bibinfo{volume}{16}},
  \bibinfo{pages}{044039} (\bibinfo{year}{2021}).

\bibitem[{\citenamefont{Zhang et~al.}(2020)\citenamefont{Zhang, Hsieh, Zhang,
  and Yao}}]{zhang2020differentiable}
\bibinfo{author}{\bibfnamefont{S.-X.} \bibnamefont{Zhang}},
  \bibinfo{author}{\bibfnamefont{C.-Y.} \bibnamefont{Hsieh}},
  \bibinfo{author}{\bibfnamefont{S.}~\bibnamefont{Zhang}}, \bibnamefont{and}
  \bibinfo{author}{\bibfnamefont{H.}~\bibnamefont{Yao}},
  \bibinfo{journal}{arXiv:2010.08561}  (\bibinfo{year}{2020}).

\bibitem[{\citenamefont{Zhang et~al.}(2021)\citenamefont{Zhang, Hsieh, Zhang,
  and Yao}}]{zhang2021neural}
\bibinfo{author}{\bibfnamefont{S.-X.} \bibnamefont{Zhang}},
  \bibinfo{author}{\bibfnamefont{C.-Y.} \bibnamefont{Hsieh}},
  \bibinfo{author}{\bibfnamefont{S.}~\bibnamefont{Zhang}}, \bibnamefont{and}
  \bibinfo{author}{\bibfnamefont{H.}~\bibnamefont{Yao}},
  \bibinfo{journal}{Mach. Learn.: Sci. Technol.}  (\bibinfo{year}{2021}).

\bibitem[{\citenamefont{He et~al.}(2021)\citenamefont{He, Li, Zheng, Li, and
  Situ}}]{he2021variational}
\bibinfo{author}{\bibfnamefont{Z.}~\bibnamefont{He}},
  \bibinfo{author}{\bibfnamefont{L.}~\bibnamefont{Li}},
  \bibinfo{author}{\bibfnamefont{S.}~\bibnamefont{Zheng}},
  \bibinfo{author}{\bibfnamefont{Y.}~\bibnamefont{Li}}, \bibnamefont{and}
  \bibinfo{author}{\bibfnamefont{H.}~\bibnamefont{Situ}}, \bibinfo{journal}{New
  J. Phys.} \textbf{\bibinfo{volume}{23}}, \bibinfo{pages}{033002}
  (\bibinfo{year}{2021}).

\bibitem[{\citenamefont{Moro et~al.}(2021)\citenamefont{Moro, Paris, Restelli,
  and Prati}}]{Moro2021quantum}
\bibinfo{author}{\bibfnamefont{L.}~\bibnamefont{Moro}},
  \bibinfo{author}{\bibfnamefont{M.~G.} \bibnamefont{Paris}},
  \bibinfo{author}{\bibfnamefont{M.}~\bibnamefont{Restelli}}, \bibnamefont{and}
  \bibinfo{author}{\bibfnamefont{E.}~\bibnamefont{Prati}},
  \bibinfo{journal}{Commun. Phys.}  (\bibinfo{year}{2021}).

\bibitem[{\citenamefont{Ostaszewski et~al.}(2021)\citenamefont{Ostaszewski,
  Trenkwalder, Masarczyk, Scerri, and Dunjko}}]{ostaszewski2021reinforcement}
\bibinfo{author}{\bibfnamefont{M.}~\bibnamefont{Ostaszewski}},
  \bibinfo{author}{\bibfnamefont{L.~M.} \bibnamefont{Trenkwalder}},
  \bibinfo{author}{\bibfnamefont{W.}~\bibnamefont{Masarczyk}},
  \bibinfo{author}{\bibfnamefont{E.}~\bibnamefont{Scerri}}, \bibnamefont{and}
  \bibinfo{author}{\bibfnamefont{V.}~\bibnamefont{Dunjko}},
  \bibinfo{journal}{arXiv:2103.16089}  (\bibinfo{year}{2021}).

\bibitem[{\citenamefont{Kuo et~al.}(2021)\citenamefont{Kuo, Fang, and
  Chen}}]{kuo2021quantum}
\bibinfo{author}{\bibfnamefont{E.-J.} \bibnamefont{Kuo}},
  \bibinfo{author}{\bibfnamefont{Y.-L.~L.} \bibnamefont{Fang}},
  \bibnamefont{and} \bibinfo{author}{\bibfnamefont{S.~Y.-C.}
  \bibnamefont{Chen}}, \bibinfo{journal}{arXiv:2104.07715}
  (\bibinfo{year}{2021}).

\bibitem[{\citenamefont{Du et~al.}(2020{\natexlab{b}})\citenamefont{Du, Huang,
  You, Hsieh, and Tao}}]{du2020quantum}
\bibinfo{author}{\bibfnamefont{Y.}~\bibnamefont{Du}},
  \bibinfo{author}{\bibfnamefont{T.}~\bibnamefont{Huang}},
  \bibinfo{author}{\bibfnamefont{S.}~\bibnamefont{You}},
  \bibinfo{author}{\bibfnamefont{M.-H.} \bibnamefont{Hsieh}}, \bibnamefont{and}
  \bibinfo{author}{\bibfnamefont{D.}~\bibnamefont{Tao}},
  \bibinfo{journal}{arXiv:2010.10217}  (\bibinfo{year}{2020}{\natexlab{b}}).

\bibitem[{\citenamefont{Chen et~al.}(2021)\citenamefont{Chen, He, Li, Zheng,
  and Situ}}]{chen2021quantum}
\bibinfo{author}{\bibfnamefont{C.}~\bibnamefont{Chen}},
  \bibinfo{author}{\bibfnamefont{Z.}~\bibnamefont{He}},
  \bibinfo{author}{\bibfnamefont{L.}~\bibnamefont{Li}},
  \bibinfo{author}{\bibfnamefont{S.}~\bibnamefont{Zheng}}, \bibnamefont{and}
  \bibinfo{author}{\bibfnamefont{H.}~\bibnamefont{Situ}},
  \bibinfo{journal}{arXiv:2106.06248}  (\bibinfo{year}{2021}).

\bibitem[{\citenamefont{Nam et~al.}(2018)\citenamefont{Nam, Ross, Su, Childs,
  and Maslov}}]{nam2018automated}
\bibinfo{author}{\bibfnamefont{Y.}~\bibnamefont{Nam}},
  \bibinfo{author}{\bibfnamefont{N.~J.} \bibnamefont{Ross}},
  \bibinfo{author}{\bibfnamefont{Y.}~\bibnamefont{Su}},
  \bibinfo{author}{\bibfnamefont{A.~M.} \bibnamefont{Childs}},
  \bibnamefont{and} \bibinfo{author}{\bibfnamefont{D.}~\bibnamefont{Maslov}},
  \bibinfo{journal}{npj Quantum Inf.} \textbf{\bibinfo{volume}{4}},
  \bibinfo{pages}{1} (\bibinfo{year}{2018}).

\bibitem[{\citenamefont{Childs et~al.}(2019)\citenamefont{Childs, Schoute, and
  Unsal}}]{childs2019circuit}
\bibinfo{author}{\bibfnamefont{A.~M.} \bibnamefont{Childs}},
  \bibinfo{author}{\bibfnamefont{E.}~\bibnamefont{Schoute}}, \bibnamefont{and}
  \bibinfo{author}{\bibfnamefont{C.~M.} \bibnamefont{Unsal}}, in
  \emph{\bibinfo{booktitle}{14th Conference on the Theory of Quantum
  Computation, Communication and Cryptography}} (\bibinfo{year}{2019}), vol.
  \bibinfo{volume}{135}, pp. \bibinfo{pages}{3:1--3:24}.

\bibitem[{\citenamefont{Wu et~al.}(2020{\natexlab{a}})\citenamefont{Wu, Davis,
  Chong, and Iancu}}]{wu2020qgo}
\bibinfo{author}{\bibfnamefont{X.-C.} \bibnamefont{Wu}},
  \bibinfo{author}{\bibfnamefont{M.~G.} \bibnamefont{Davis}},
  \bibinfo{author}{\bibfnamefont{F.~T.} \bibnamefont{Chong}}, \bibnamefont{and}
  \bibinfo{author}{\bibfnamefont{C.}~\bibnamefont{Iancu}},
  \bibinfo{journal}{arXiv:2012.09835}  (\bibinfo{year}{2020}{\natexlab{a}}).

\bibitem[{\citenamefont{Hamilton et~al.}(2017)\citenamefont{Hamilton, Ying, and
  Leskovec}}]{hamilton2017representation}
\bibinfo{author}{\bibfnamefont{W.~L.} \bibnamefont{Hamilton}},
  \bibinfo{author}{\bibfnamefont{R.}~\bibnamefont{Ying}}, \bibnamefont{and}
  \bibinfo{author}{\bibfnamefont{J.}~\bibnamefont{Leskovec}},
  \bibinfo{journal}{arXiv:1709.05584}  (\bibinfo{year}{2017}).

\bibitem[{\citenamefont{Wu et~al.}(2020{\natexlab{b}})\citenamefont{Wu, Pan,
  Chen, Long, Zhang, and Philip}}]{wu2020comprehensive}
\bibinfo{author}{\bibfnamefont{Z.}~\bibnamefont{Wu}},
  \bibinfo{author}{\bibfnamefont{S.}~\bibnamefont{Pan}},
  \bibinfo{author}{\bibfnamefont{F.}~\bibnamefont{Chen}},
  \bibinfo{author}{\bibfnamefont{G.}~\bibnamefont{Long}},
  \bibinfo{author}{\bibfnamefont{C.}~\bibnamefont{Zhang}}, \bibnamefont{and}
  \bibinfo{author}{\bibfnamefont{S.~Y.} \bibnamefont{Philip}},
  \bibinfo{journal}{IEEE T. Neur. Net. Lear.} \textbf{\bibinfo{volume}{32}},
  \bibinfo{pages}{4} (\bibinfo{year}{2020}{\natexlab{b}}).

\bibitem[{\citenamefont{Kingma and Welling}(2013)}]{kingma2013auto}
\bibinfo{author}{\bibfnamefont{D.~P.} \bibnamefont{Kingma}} \bibnamefont{and}
  \bibinfo{author}{\bibfnamefont{M.}~\bibnamefont{Welling}},
  \bibinfo{journal}{arXiv:1312.6114}  (\bibinfo{year}{2013}).

\bibitem[{\citenamefont{Jin et~al.}(2018)\citenamefont{Jin, Barzilay, and
  Jaakkola}}]{jin2018junction}
\bibinfo{author}{\bibfnamefont{W.}~\bibnamefont{Jin}},
  \bibinfo{author}{\bibfnamefont{R.}~\bibnamefont{Barzilay}}, \bibnamefont{and}
  \bibinfo{author}{\bibfnamefont{T.}~\bibnamefont{Jaakkola}}, in
  \emph{\bibinfo{booktitle}{International Conference on Machine Learning}}
  (\bibinfo{year}{2018}), pp. \bibinfo{pages}{2323--2332}.

\bibitem[{\citenamefont{Chung et~al.}(2014)\citenamefont{Chung, Gulcehre, Cho,
  and Bengio}}]{69e088c8129341ac89810907fe6b1bfe}
\bibinfo{author}{\bibfnamefont{J.}~\bibnamefont{Chung}},
  \bibinfo{author}{\bibfnamefont{C.}~\bibnamefont{Gulcehre}},
  \bibinfo{author}{\bibfnamefont{K.}~\bibnamefont{Cho}}, \bibnamefont{and}
  \bibinfo{author}{\bibfnamefont{Y.}~\bibnamefont{Bengio}},
  \bibinfo{journal}{arXiv:1412.3555}  (\bibinfo{year}{2014}).

\bibitem[{Asp()}]{Aspen11QPU}
\bibinfo{howpublished}{\url{https://qcs.rigetti.com/qpus/}}.

\bibitem[{\citenamefont{Fan et~al.}(2018)\citenamefont{Fan, Lewis, and
  Dauphin}}]{fan2018hierarchical}
\bibinfo{author}{\bibfnamefont{A.}~\bibnamefont{Fan}},
  \bibinfo{author}{\bibfnamefont{M.}~\bibnamefont{Lewis}}, \bibnamefont{and}
  \bibinfo{author}{\bibfnamefont{Y.}~\bibnamefont{Dauphin}}, in
  \emph{\bibinfo{booktitle}{In Proceedings of the 56th Annual Meeting of the
  Association for Computational Linguistics (ACL)}} (\bibinfo{year}{2018}), pp.
  \bibinfo{pages}{889--898}.

\bibitem[{\citenamefont{Bergholm et~al.}(2018)\citenamefont{Bergholm, Izaac,
  Schuld, Gogolin, Alam, Ahmed, Arrazola, Blank, Delgado, Jahangiri
  et~al.}}]{bergholm2018pennylane}
\bibinfo{author}{\bibfnamefont{V.}~\bibnamefont{Bergholm}},
  \bibinfo{author}{\bibfnamefont{J.}~\bibnamefont{Izaac}},
  \bibinfo{author}{\bibfnamefont{M.}~\bibnamefont{Schuld}},
  \bibinfo{author}{\bibfnamefont{C.}~\bibnamefont{Gogolin}},
  \bibinfo{author}{\bibfnamefont{M.~S.} \bibnamefont{Alam}},
  \bibinfo{author}{\bibfnamefont{S.}~\bibnamefont{Ahmed}},
  \bibinfo{author}{\bibfnamefont{J.~M.} \bibnamefont{Arrazola}},
  \bibinfo{author}{\bibfnamefont{C.}~\bibnamefont{Blank}},
  \bibinfo{author}{\bibfnamefont{A.}~\bibnamefont{Delgado}},
  \bibinfo{author}{\bibfnamefont{S.}~\bibnamefont{Jahangiri}},
  \bibnamefont{et~al.}, \bibinfo{journal}{arXiv:1811.04968}
  (\bibinfo{year}{2018}).

\bibitem[{\citenamefont{Kingma and Ba}(2014)}]{kingma2014adam}
\bibinfo{author}{\bibfnamefont{D.~P.} \bibnamefont{Kingma}} \bibnamefont{and}
  \bibinfo{author}{\bibfnamefont{J.}~\bibnamefont{Ba}},
  \bibinfo{journal}{arXiv:1412.6980}  (\bibinfo{year}{2014}).

\bibitem[{\citenamefont{Zhang et~al.}(2019)\citenamefont{Zhang, Jiang, Cui,
  Garnett, and Chen}}]{zhang2019d}
\bibinfo{author}{\bibfnamefont{M.}~\bibnamefont{Zhang}},
  \bibinfo{author}{\bibfnamefont{S.}~\bibnamefont{Jiang}},
  \bibinfo{author}{\bibfnamefont{Z.}~\bibnamefont{Cui}},
  \bibinfo{author}{\bibfnamefont{R.}~\bibnamefont{Garnett}}, \bibnamefont{and}
  \bibinfo{author}{\bibfnamefont{Y.}~\bibnamefont{Chen}}, in
  \emph{\bibinfo{booktitle}{Proceedings of the 33rd International Conference on
  Neural Information Processing Systems}} (\bibinfo{year}{2019}), pp.
  \bibinfo{pages}{1588--1600}.

\end{thebibliography}

\clearpage

\appendix
\section{Hyperparameters in simulations}
We implement the simulations on a classical computer with a CPU i9-10900K using Pennylane\cite{bergholm2018pennylane}, which includes a wide range of quantum machine learning libraries.
The generator and the predictor are trained with Adam optimizer\cite{kingma2014adam}. The hyperparameters of the proposed method and DQAS are shown in Table \ref{tab:Hyperparameters0} and Table \ref{tab:Hyperparameters of dqas}.

\begin{table}[htbp]
	\centering
	\caption{The hyperparameters of the our method.}
	\begin{tabular}{ccc}
		\toprule
		\hline
		\multicolumn{2}{c}{Hyperparameters} & Value \\
		\midrule
		\hline
		\multirow{10}[2]{*}{} & Dimenstion of $\mathbf{h}_v$& 56 \\
		& Dimenstion of $\mathbf{z}$ & 56 \\
		& Batch size of meta-training  & 32 \\
		& Learning rates for training the generator and predictor  & 1e-4 \\
		& Weighted parameter $\lambda$ & 1e-5 \\
		& Number of training epochs for the generator & 400 \\
		& Number of training epochs for the predictor & 100 \\
		\bottomrule
		\hline
	\end{tabular}%
	\label{tab:Hyperparameters0}%
\end{table}%
\label{DQAS superparameters}

\begin{table}[htbp]
	\centering
	\caption{The hyperparameters of DQAS.}
	\begin{tabular}{cc}
		\toprule
		\hline
		Hyperparameter & Value \\
		\hline
		\midrule
		Batch size  & 256 \\
		Number of training epochs  & 300 \\
		Learning rate of the structure & 0.2 \\
		Learning rate of the gate parameters & 0.1 \\
		\bottomrule
		\hline
	\end{tabular}%
	\label{tab:Hyperparameters of dqas}%
\end{table}%

\section{Details of graph computation}

In order to extract the structure information  from the DAG $\mathcal{G} $ more accurately, we use a bidirectional encoding strategy \cite{zhang2019d}, which computes the graph state in both the forward and backward directions.
Specifically, we use two GRUs to obtain the forward  and the backward graph states ($ \boldsymbol{h}_f$ and $\boldsymbol{h}_b$), respectively.
We concatenate $ \boldsymbol{h}_f $ and $ \boldsymbol{h}_b $ and use a trainable single linear layer to reduce the dimension of the concatenated vector by half.
We use this vector to denote the final graph state $ \boldsymbol{h}_{\mathcal{G}} $.
The position information of the previously connected nodes is also considered in the calculation of $ \mathbf{h}_{v} $
\begin{equation}
	\begin{aligned}
		\mathbf{h}_{v}^{\text {in}}=\sum_{u \rightarrow v} g\left(\mbox{Concat}(\mathbf{h}_{u},\mathbf{x}_{\text{pos}})\right) \odot m\left(\mbox{Concat}(\mathbf{h}_{u},\mathbf{x}_{\text{pos}})\right),
	\end{aligned}
\end{equation}
where $ \mathbf{x}_{\text{pos}} $ is a one-hot vector of the node $ u $'s global position.

\section{Training dataset of the generator}
\label{sec:generator}
Each training sample of the generator consists of a target circuit and the optimal compiled one with the native gates.
The training set consists of target circuits with different lengths, \ie $L=4, 5, 6$.
We randomly generate 1000 target circuits for each length , leading to 3000 target circuits in the training set.
In this paper, we use the DQAS algorithm \cite{zhang2020differentiable} to search the optimal compiled circuits of the target circuits.
It should be noted that any other QAS algorithm can be used to search the optimal compiled circuits.
We search the compiled circuit for each target circuit by gradually increasing the length of the compiled circuit until the loss is lower than 0.05 or the circuit length reaches $5L$.
If the loss of the compiled circuit with $5L$ gates is higher than 0.05, we will give up the corresponding target circuit as we cannot find a compiled circuit within a limited length.


\section{Training and test dataset of the predictor}
\label{sec:AppendixE}
Each training sample of the predictor consists of a target circuit $ \mathcal{G}_{t_i} $, a compiled circuit $ \mathcal{G}_{c_i} $ and the corresponding loss $s_i$.
The training set of the predictor should include the compiled circuits with a variety of performances.
We use the 3,000 training samples of the generator and calculate the corresponding losses.
In addition, we also collect 207,000 target circuits which are generated by randomly selecting 4, 5 and 6 gates from  $\mathcal{A}_{\mathit{target}}$ and randomly choosing the operated qubits. Their compiled circuits are generated by randomly selecting $[2L, 5L]$ gates from $\mathcal{A}_{\mathit{native}} $ and randomly choosing the operated qubits, where $L$ is the number of quantum gates in the target circuit.
The 210,000 samples are randomly divided into the training and test sets of the predictor, which consist of 200,000 and 10,000 samples, respectively.

\end{document}